\documentclass[preprint,aps,showpacs,prb]{revtex4}

\usepackage{graphicx}% Include figure files
\usepackage{dcolumn}% Align table columns on decimal point
\usepackage{bm}% bold math
\usepackage{epsfig}
\graphicspath{{./}{data_paper/}}
\DeclareGraphicsExtensions{.ps,.ps.gz}
\begin{document}

\title{\bf Sources of negative differential resistance in electric nanotransport}

\author{Ioan B\^aldea}
 \altaffiliation[Also at ]{National Institute for Lasers, Plasma, and Radiation Physics, 
ISS, RO 077125, Bucharest, Romania}
\author{Horst K\"oppel}%
\affiliation{%
Theoretische Chemie, Universit\"at Heidelberg, Im Neuenheimer Feld 229, 
D-69120 Heidelberg, Germany}

% \date{\today}

\begin{abstract}
A negative differential resistance (NDR) in nanotransport is often 
ascribed to electron correlations. We present a simple example revealing that 
finite electrode bandwidths and energy dependent electrode density of states 
can cause a significant NDR, which may occur even in uncorrelated systems.
So, special care is needed in assessing the role of electron correlations 
in the NDR.
\end{abstract}

% 73.63.-b	Electronic transport in nanoscale materials and structures
% 73.63.Kv	Quantum dots

% 85.35.-p	Nanoelectronic devices
% 85.35.Be	Quantum well devices (quantum dots, quantum wires, etc.)
% 85.35.Gv	Single electron devices

% 85.65.+h	Molecular electronic devices

\maketitle
%%%%%%%%%%%%%%%%%%%%%%%%%%%%%%%%%%%%%%%%%%%%%%%%%%%%%%%%%%%%%%%%%%%%%%
The fact that the current-voltage ($I$-$V$) characteristics of the dc-transport 
can exhibit a negative differential resistance (NDR) in systems described 
within a single-particle picture, and is not necessarily related to 
electron correlations is well known in semiconductor physics.\cite{Frensley:91} 
However, in the nanophysics community 
the NDR in the $I$-$V$ curve is often ascribed to (presumably strong) electron correlations. 
In fact, some calculations performed on simple but nontrivial models of correlated 
electrons, like the interacting resonant level model, found no NDR effect far away from 
resonance,\cite{Mehta:06,Mehta:07}
while other calculations revealed a more \cite{Schmitteckert:08} or less \cite{Doyon:07,Nishino:09} 
pronounced NDR effect at resonance.
At the end of this note, we shall return to the NDR effect within the interacting resonant model.
Beforehand --- and this is the main aim of the present work --- 
we want to emphasize that other, more common sources of 
the NDR are relevant for nanotransport as well. Therefore, special care is needed if one attempts 
to ascribe the NDR to electron correlations. 

The naive ``argument'' behind the confusion that the NDR is an electron correlation effect 
seems to be the following. Within the Landauer approach of the transport in uncorrelated systems,
the current resulting from the imbalance 
between the source and drain chemical potentials $\mu_S = \varepsilon_F + eV_{sd}/2$ and 
$\mu_D = \varepsilon_F - eV_{sd}/2$ 
is expressed as an integral of the transmission coefficient $T(\varepsilon)$
over energies from $\varepsilon = \mu_D$  to $\varepsilon = \mu_S$. An NDR cannot occur because
the current monotonically increases, since the integrand is positive 
[$T(\varepsilon)\geq 0$] and the integration range increases as the voltage $V_{sd}$
becomes higher.

\begin{figure}[htb]
% $ $\\[7ex]
\centerline{\hspace*{-0ex}\includegraphics[width=0.5\textwidth,angle=0]{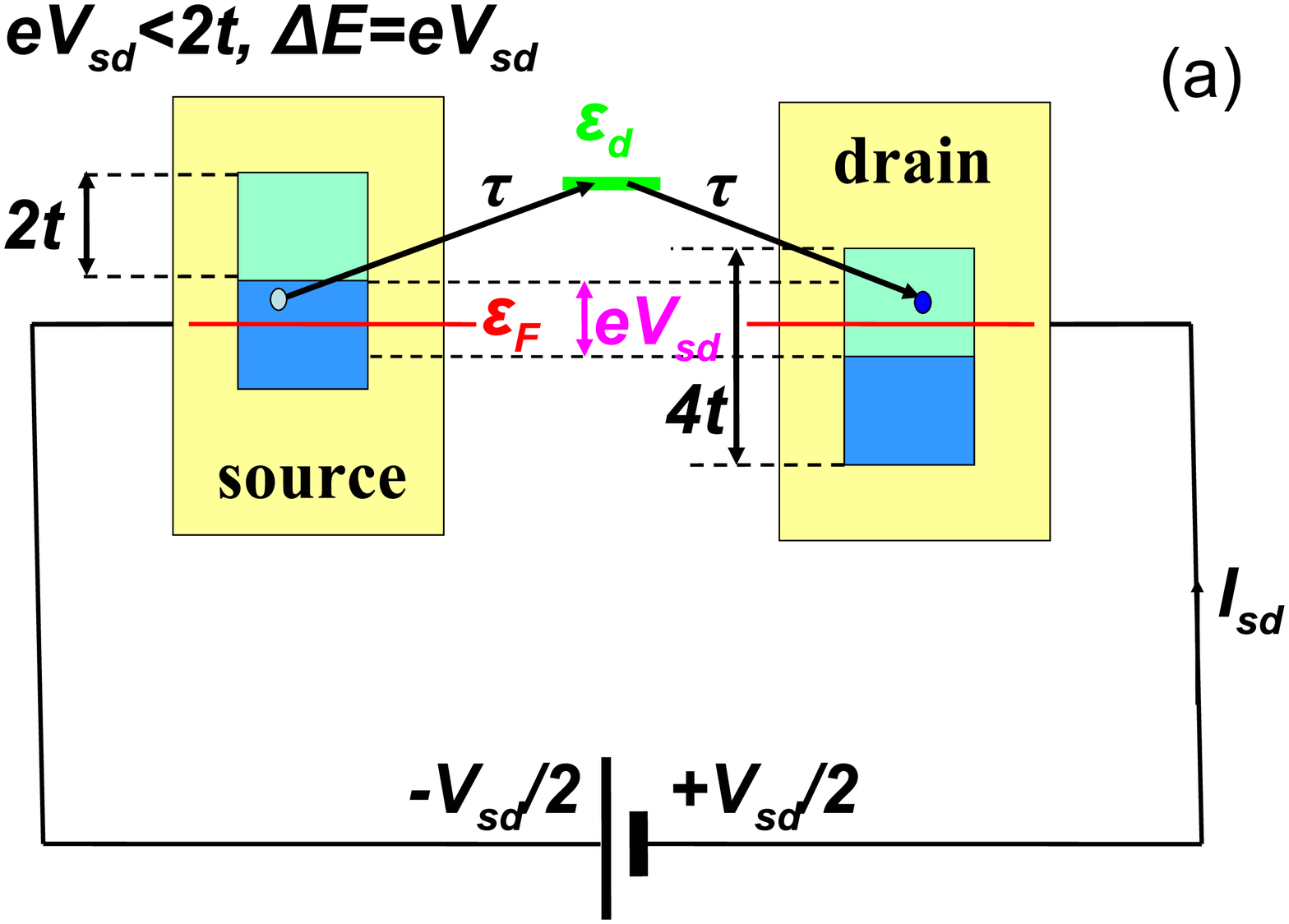}}
\centerline{\hspace*{-0ex}\includegraphics[width=0.5\textwidth,angle=0]{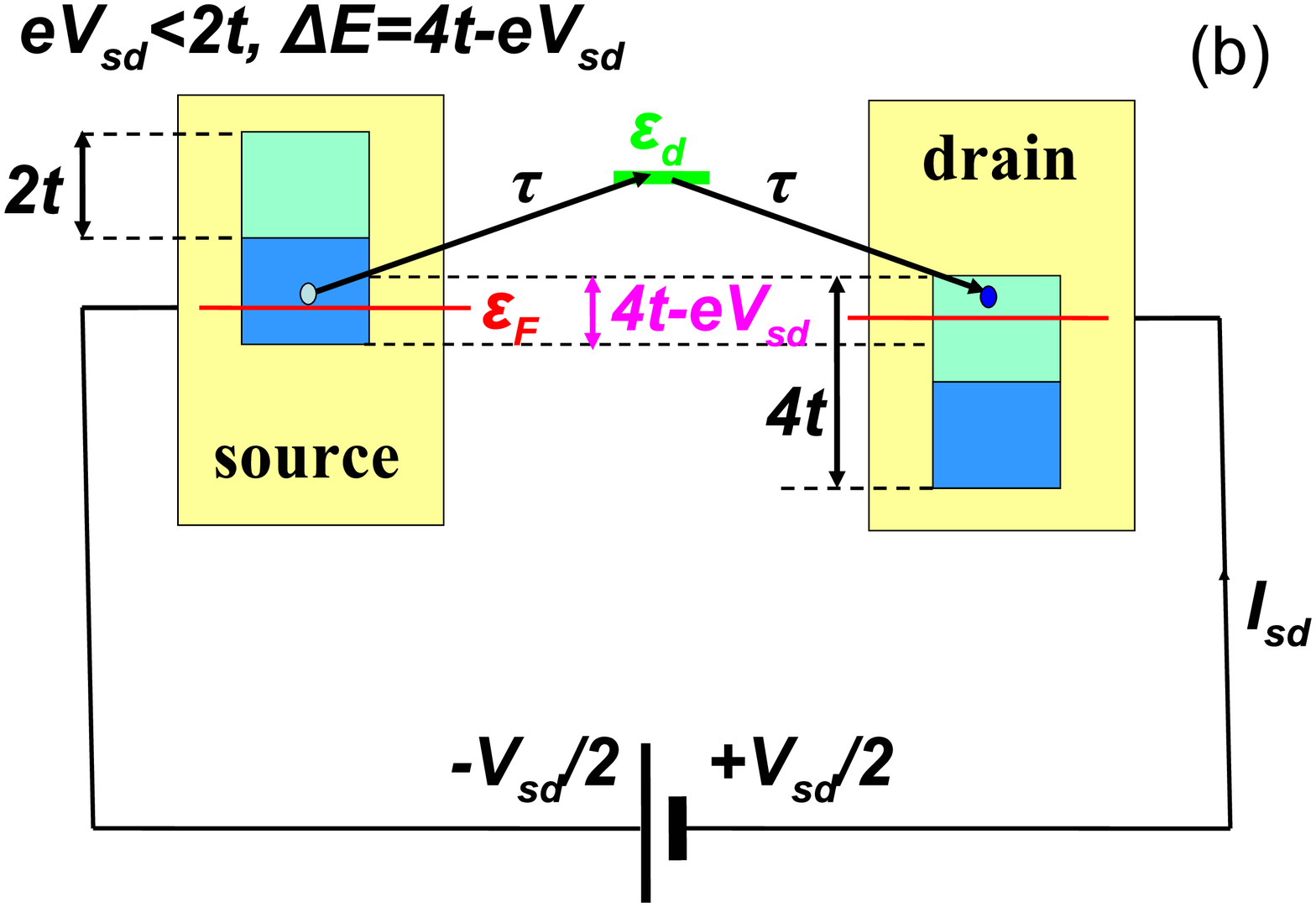}}
\caption{\label{fig:setup} (Color online) 
Schematical representation of a typical two-terminal setup. 
By gradually increasing the source-drain voltage $V_{sd}$ 
the energy window $\Delta E$ of the allowed elastic 
tunneling processes (a) increases for $ e V_{sd} < 2 t$, 
but (b) beyond the point 
$e V_{sd} = 2t$ (electrode half-bandwidth) it decreases. 
Elastic tunneling cannot occur for $e V_{sd} \geq 4 t$.}
\end{figure}

To illustrate that this is not the case, let us consider a two-terminal setup 
(Fig.~\ref{fig:setup}), consisting of a nanosystem [quantum dot(s) or molecule(s)]
linked to semi-infinite leads (source and drain) at zero temperature. 
For simplicity, their 
bandwidth $4t$ as well as their coupling to (say,) the dot $\tau$ will be supposed 
to be identical. By gradually rising
the source-drain voltage $V_{sd}$ starting from $V_{sd}=0$, the drain current $I_{sd}$ 
will first progressively 
increase because the energy window $\Delta E$ of the (elastic) electron tunneling processes allowed 
by Pauli's principle 
becomes broader (Fig.~\ref{fig:setup}a). However, further increasing $V_{sd}$ beyond 
half of the electrode bandwidth ($e V_{sd}^{\ast} \equiv 2 t$) will diminish this energy window 
(Fig.~\ref{fig:setup}b),
and this will be accompanied by a current reduction, which becomes more and more pronounced 
as the electrode band edge is approach. For $e V_{sd} \geq 4t$, elastic tunneling is no longer 
possible, and the current is completely blocked ($I_{sd}=0$). 
This fact that the current $I_{sd}$ should diminish as $V_{sd}$ 
exceeds $V_{sd}^{\ast}$ and is completely suppressed 
above  the band edge ($4 t$) applies for a general two-terminal setup for a 
sufficiently weak hybridization $\Gamma_0 \equiv 2 \tau^2/t$.
%%%%%%%%%%%%%%%%%%%%%%%%%%%%%%%%%%%%%%%%%%%%%%%%%%%%%%%%%%%%%%%%%%%%%%%
%%%%%%%%%%%%%%%%%%%%%%%%%%%%%%%%%%%%%%%%%%%%%%%%%%%%%%%%%%%%%%%%%%%%%%

To make the analysis more specific, let us consider 
a point contact (noninteracting resonant level) 
model, wherein the nanosystem consists of a single nondegenerate energy level $\varepsilon_g$ 
linked to one-dimensional semi-infinite electrodes. The second-quantized Hamiltonian reads
%%%%%%%%%%%%%%%%%%%%%%%%%%%%%%%%%%%%%%%%%%%%%%%%%%%%%%%%%%%%%%%%%%%%%%%%%%%%%%%%%%%
\begin{eqnarray}
H & = & -t \sum_{l\leq -1} \left (c_{l}^{\dagger} c_{l-1} + h.c.\right) +
\mu_S \sum_{l\leq -1} c_{l}^{\dagger} c_{l} \nonumber \\
 & & -t \sum_{l\geq 1} \left (c_{l}^{\dagger} c_{l+1} + h.c.\right) + 
\mu_D \sum_{l\geq 1} c_{l}^{\dagger} c_{l} \label{eq-ham}\\
& & + \varepsilon_g c_{0}^\dagger c_{0} - 
\tau \left( c_{-1}^\dagger c_{0} + c_{1}^\dagger c_{0} + h.c. \right) \ . \nonumber
\end{eqnarray}
%%%%%%%%%%%%%%%%%%%%%%%%%%%%%%%%%%%%%%%%%%%%%%%%%%%%%%%%%%%%%%%%%%%%%%%%%%%%%%%%%%%
As usual, we set $t = 1$ and $\varepsilon_F = 0$. We assume
$\varepsilon_g \geq 0$ (n-type conduction)
for simplicity, but because model (\ref{eq-ham}) possesses
particle-hole symmetry, one can replace 
$\varepsilon_g$ by $\vert\varepsilon_g\vert$ below.
The electrode-dot coupling $\tau$ yields well known expressions of the 
embedding self-energies 
$
\Sigma_x(\varepsilon) = \Delta_x(\varepsilon) - i\Gamma_x(\varepsilon)/2 
$
($x=S,D$), where \cite{Caroli:71,Nitzan:01}
%%%%%%%%%%%%%%%%%%%%%%%%%%%%%%%%%%%%%%%%%%%%%%%%%%%%%%%%%%%%%%%%%%%%%%%%%%%%%%%%%%%
\begin{eqnarray}
\displaystyle
\Delta_x(\varepsilon) & = & \Delta(\varepsilon - \mu_x); 
\Gamma_x(\varepsilon) = \Gamma(\varepsilon - \mu_x); 
\nonumber \\
\Delta(\varepsilon) & = & \frac{\tau^2\varepsilon}{2 t^2}; 
\Gamma(\varepsilon) = \frac{\tau^2}{t^2}
\sqrt{4 t^2 - \varepsilon^2} \ \theta(2 t - \vert \varepsilon\vert) .
\label{eq-Sigma-x}
\end{eqnarray}
%%%%%%%%%%%%%%%%%%%%%%%%%%%%%%%%%%%%%%%%%%%%%%%%%%%%%%%%%%%%%%%%%%%%%%%%%%%%%%%%%%%
They can be inserted into the Dyson equation 
%%%%%%%%%%%%%%%%%%%%%%%%%%%%%%%%%%%%%%%%%%%%%%%%%%%%%%%%%%%%%%%%%%%%%%%%%%%%%%%%%%%
\begin{equation}
G^{-1}(\varepsilon) = \varepsilon - \varepsilon_g - \Sigma_S (\varepsilon) - \Sigma_D(\varepsilon)
\end{equation}
%%%%%%%%%%%%%%%%%%%%%%%%%%%%%%%%%%%%%%%%%%%%%%%%%%%%%%%%%%%%%%%%%%%%%%%%%%%%%%%%%%%
to obtain the retarded Green function $G(\varepsilon)$ of the embedded dot.  
With the aid of the latter, the electric current can be expressed as 
(electron spin is disregarded)
%%%%%%%%%%%%%%%%%%%%%%%%%%%%%%%%%%%%%%%%%%%%%%%%%%%%%%%%%%%%%%%%%%%%%%%%%%%%%%%%%%%
\begin{eqnarray}
\displaystyle
I_{sd} & = & \frac{e}{h}\int_{\mu_D}^{\mu_S} d\,\varepsilon T(\varepsilon) = 
\frac{e}{h}\int_{\mu_D}^{\mu_S} d\,\varepsilon 
\Gamma_S(\varepsilon) \Gamma_D(\varepsilon) \vert G(\varepsilon)\vert^2 , \nonumber \\
& = & \frac{e}{h}\int_{\mu_D}^{\mu_S} d\,\varepsilon 
\frac{\Gamma_D(\varepsilon) \Gamma_S(\varepsilon)
}
{
\left[\varepsilon - \varepsilon_g - \overline{\Delta}(\varepsilon))\right]^2
+ \overline{\Gamma}(\varepsilon)^2/4
} \ ,
\label{eq-I}
\end{eqnarray}
%%%%%%%%%%%%%%%%%%%%%%%%%%%%%%%%%%%%%%%%%%%%%%%%%%%%%%%%%%%%%%%%%%%%%%%%%%%%%%%%%%%
where 
$\overline{\Gamma}(\varepsilon) \equiv \Gamma_D(\varepsilon) + \Gamma_S(\varepsilon)$
and 
$\overline{\Delta}(\varepsilon) \equiv \Delta_D(\varepsilon) + \Delta_S(\varepsilon)$.

\begin{figure}[htb]
% $ $\\[7ex]
% \centerline{\hspace*{-0ex}\includegraphics[width=0.5\textwidth,angle=0]{fig_various_td_ed_0.eps}}
\centerline{\hspace*{-0ex}\includegraphics[width=0.5\textwidth,angle=0]{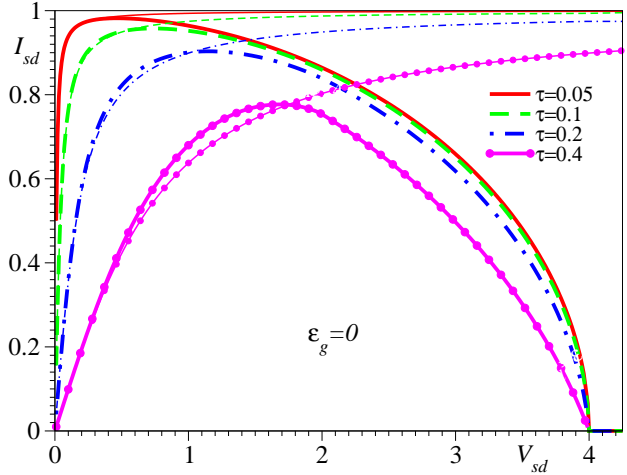}}
\caption{\label{fig:on} (Color online) $I$-$V$ curves at resonance 
($\varepsilon_g =\varepsilon_F = 0$) for $\tau = 0.05; 0.1; 0.2; 0.4$ computed exactly 
(thick lines) and within approximation (i) described in the text (thin lines).
Current $I_{sd}$ in units $ I_{sd}^{s} = \pi e\Gamma_0/h$.}
\end{figure}

\begin{figure}[htb]
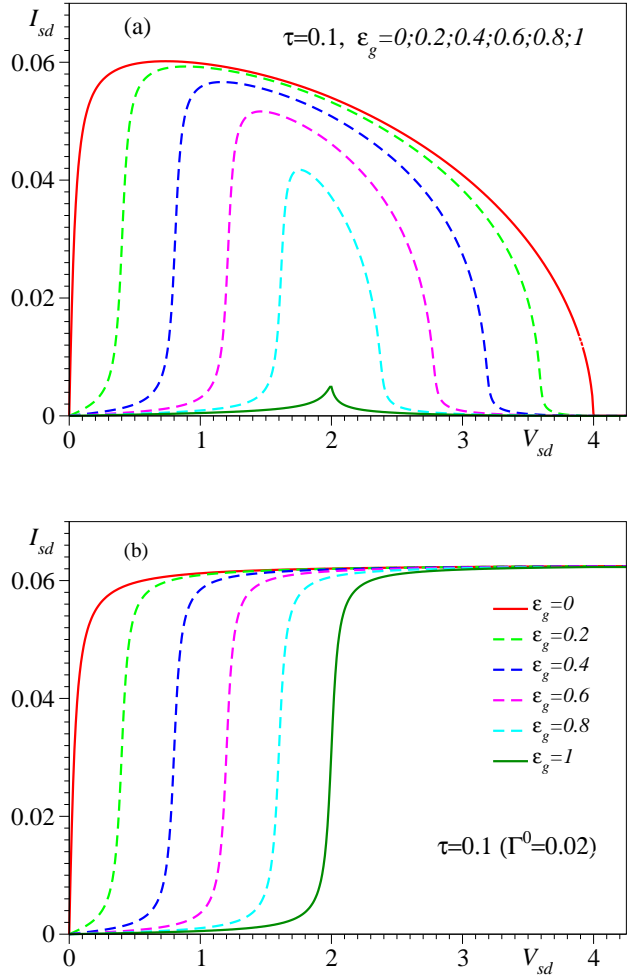

% $ $\\[7ex]
% \centerline{\hspace*{-0ex}\includegraphics[width=0.3\textwidth,angle=0]{fig_td_0.1_various_ed_exact.eps}}
\centerline{\hspace*{-0ex}\includegraphics[width=0.5\textwidth,angle=0]{fig3a.eps}}
$  $\\[0.2ex]
% \centerline{\hspace*{-0ex}\includegraphics[width=0.3\textwidth,angle=0]{fig_td_0.1_various_ed_wbl.eps}}
\centerline{\hspace*{-0ex}\includegraphics[width=0.5\textwidth,angle=0]{fig3b.eps}}
\caption{\label{fig:off}(Color online)
$I$-$V$ curves out of resonance 
for $\tau = 0.1$ ($\Gamma_0=0.02$) 
computed (a) exactly and (b) within approximation (i) described in the text 
for 
$\varepsilon_g =0; 0.2; 0.4; 0.6; 0.8; 1$
(values increasing downwards). Current in units $et/h$.}
\end{figure}

$I$-$V$ characteristics computed exactly by means of Eq.~(\ref{eq-I}) 
at resonance ($\varepsilon_g = 0$) are depicted by the thick lines in Fig.~\ref{fig:on}.
These curves show that, indeed, the current is suppressed as the bias
approaches the bandwidth and disappears beyond $e V_{sd} > 4t$.
Away from resonance ($\varepsilon_g \neq 0$), 
a new aspect is visible in Fig.~\ref{fig:off}a. The current vanishes 
even below the bandwidth $4 t$. Practically, the suppression is complete at 
$V_{sd} = 4 t - \varepsilon_g$; beyond this value, 
the $I$-$V$ curves only exhibit negligible tails of widths $\sim \Gamma_0 = 2\tau^2/t$.
On the other side, the exact $I$-$V$ characteristics of Figs.~\ref{fig:on} and \ref{fig:off}a
reveal that the current decreases well before reaching the value $V_{sd}^{\ast}=2 t/e$, 
which one could expect from Fig.~\ref{fig:setup}.
This demonstrates that the finite bandwidth effect discussed above is only \emph{one} reason 
why the NDR should occur. 
%%%%%%%%%%%%%%%%%%%%%%%%%%%%%%%%%%%%%%%%%%%%%%%%%%%%%%%%%%%%%%%%%%%%%%%
%%%%%%%%%%%%%%%%%%%%%%%%%%%%%%%%%%%%%%%%%%%%%%%%%%%%%%%%%%%%%%%%%%%%%%
%%%%%%%%%%%%%%%%%%%%%%%%%%%%%%%%%%%%%%%%%%%%%%%%%%%%%%%%%%%%%%%%%%%%%%%
%%%%%%%%%%%%%%%%%%%%%%%%%%%%%%%%%%%%%%%%%%%%%%%%%%%%%%%%%%%%%%%%%%%%%%

Significant physical insight can be gained by examining three limits of Eq.~(\ref{eq-I}):

(i) One can approximate the embedding energies by their values at 
$\varepsilon = \mu_x$ ($\Sigma_{S,D} \simeq -i\Gamma_0/2$) in the 
\emph{whole} integration range, which means to simply ignore the 
$\theta$ step functions in Eq.~(\ref{eq-Sigma-x}). One then gets the current
%%%%%%%%%%%%%%%%%%%%%%%%%%%%%%%%%%%%%%%%%%%%%%%%%%%%%%%%%%%%%%%%%%%%%%
\begin{equation}
\displaystyle
I_{sd}^{low} = \frac{e\Gamma_0}{h} % \frac{e}{h} \Gamma_0
\left(
\arctan\frac{e V -  2\varepsilon_g}{2\Gamma_0} + 
\arctan\frac{e V + 2\varepsilon_g}{2\Gamma_0}
\right) .
\label{eq-wbl}
\end{equation}
%%%%%%%%%%%%%%%%%%%%%%%%%%%%%%%%%%%%%%%%%%%%%%%%%%%%%%%%%%%%%%%%%%%%%%
As this amounts to assume that 
the electrode bandwidth is the largest energy scale
(more precisely, for $ V_{sd}, \varepsilon_g, \tau \ll t$), 
Eq.~(\ref{eq-wbl}) is usually referred to as the wide band limit.

(ii) Next, one can compute the current using the electrode density of states (DOS) 
$\Gamma_x$ for $\varepsilon = \mu_x$, but unlike above, considering the Heaviside 
$\theta$ functions in Eq.~(\ref{eq-Sigma-x}) 
%%%%%%%%%%%%%%%%%%%%%%%%%%%%%%%%%%%%%%%%%%%%%%%%%%%%%%%%%%%%%%%%%%%%%%
\begin{equation}
\displaystyle
I_{sd}^{fb} = \frac{e \Gamma_0}{h\left(1-\tau^2/t^2\right)}
\left(
\arctan\frac{\Lambda_{+}}{2\Gamma_0} + 
\arctan\frac{\Lambda_{-}}{2\Gamma_0}
\right) ,
\label{eq-fb}
\end{equation}
%%%%%%%%%%%%%%%%%%%%%%%%%%%%%%%%%%%%%%%%%%%%%%%%%%%%%%%%%%%%%%%%%%%%%%
where 
$\Lambda_{\pm} \equiv \left[\min(e V_{sd}, 4 t - e V_{sd}) \pm 2 \varepsilon_g\right]
\times \left(1 - \tau^2/t^2\right)$.
Similar to approximation (i), the electrode DOS is 
assumed constant, but the fact that the electrode bandwidths are \emph{finite} 
(the main physical aspect underlying Fig.~\ref{fig:setup})
is taken into account by this approximation.  

\begin{figure}[htb]
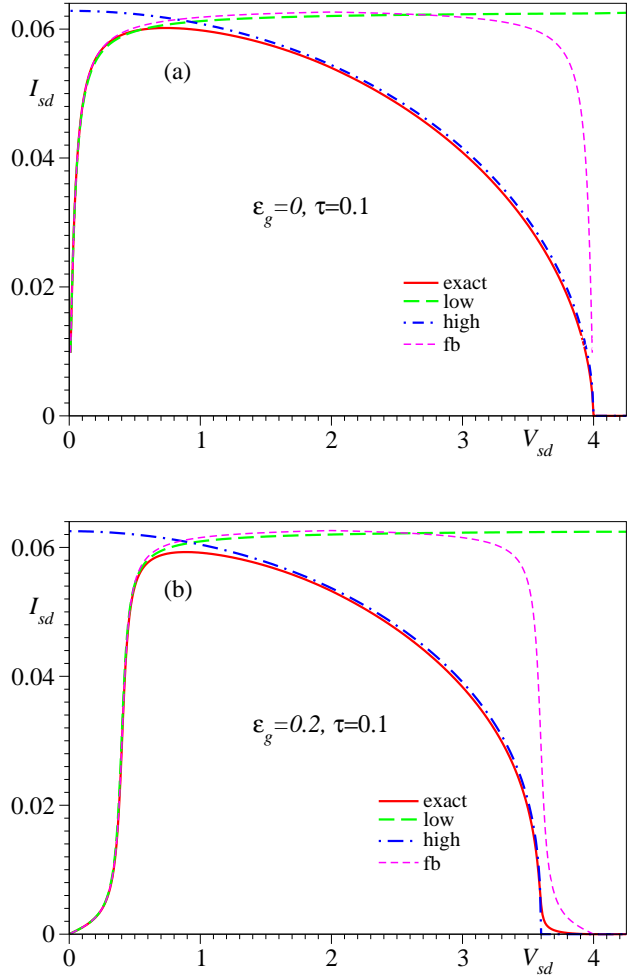

% $ $\\[7ex]
% \centerline{\hspace*{-0ex}\includegraphics[width=0.3\textwidth,angle=0]{fig_td_0.1_ed_0.eps}}
\centerline{\hspace*{-0ex}\includegraphics[width=0.5\textwidth,angle=0]{fig4a.eps}}
$  $\\[0.2ex]
% \centerline{\hspace*{-0ex}\includegraphics[width=0.3\textwidth,angle=0]{fig_td_0.1_ed_0.2.eps}}
\centerline{\hspace*{-0ex}\includegraphics[width=0.5\textwidth,angle=0]{fig4b.eps}}
\caption{\label{fig:exact-vs-approx} (Color online)
$I$-$V$ curves for $\tau = 0.1$ computed 
exactly and within the approximations described in the text: (a) at resonance 
$\varepsilon_d = 0$ and (b) out of resonance, $\varepsilon_d = 0.2$. 
Current in units $et/h$. Labels as in Eqs.~(\ref{eq-wbl}), (\ref{eq-fb}), and (\ref{eq-be}).}
\end{figure}

\begin{figure}[htb]
% $ $\\[7ex]
% \centerline{\hspace*{-0ex}\includegraphics[width=0.3\textwidth,angle=0]{fig_ed_0_various_td_fb.eps}}
\centerline{\hspace*{-0ex}\includegraphics[width=0.5\textwidth,angle=0]{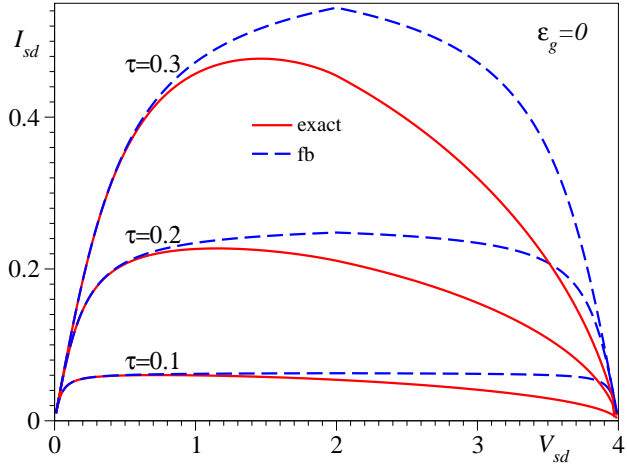}}
\caption{\label{fig:fb} (Color online)
$I$-$V$ curves on resonance ($\varepsilon_g = 0$) 
for the three electrode-dot couplings $\tau$ specified in the inset 
computed exactly and within approximation (ii) described in the 
text (label $fb$). Notice that the latter exhibit a cusp at 
$ e V_{sd} = 2 t$ that marks the NDR onset in this approximation, 
which can be substantially higher 
than the exact NDR onset.}
\end{figure}

(iii) 
Because the main contribution to the integral in Eq.~(\ref{eq-I}) 
comes from the pole of the 
Green function of the isolated dot, one can use the embedding 
energies calculated at $\varepsilon = \varepsilon_g$. In fact, this 
approximation yields very accurate $I$-$V$ curves, which are not shown 
because they could be hardly distinguished from the exact curves within the drawing 
accuracy of Figs.~\ref{fig:on}, \ref{fig:off}a, \ref{fig:exact-vs-approx}, and \ref{fig:fb}.
More instructive is however to furthermore assume 
that the voltage $V_{sd}$ is sufficiently high and extend the integration in Eq.~(\ref{eq-I}) 
from $-\infty$ to $+\infty$. The result is
%%%%%%%%%%%%%%%%%%%%%%%%%%%%%%%%%%%%%%%%%%%%%%%%%%%%%%%%%%%%%%%%%%%%%%%
\begin{equation}
\displaystyle
I_{sd}^{high} = \frac{e}{\hbar} 
\frac{\Gamma(\varepsilon_g - eV/2) \Gamma(\varepsilon_g + eV/2)}
{\Gamma(\varepsilon_g - eV/2) + \Gamma(\varepsilon_g + eV/2)} .
\label{eq-be}
\end{equation}
%%%%%%%%%%%%%%%%%%%%%%%%%%%%%%%%%%%%%%%%%%%%%%%%%%%%%%%%%%%%%%%%%%%%%%%

$I$-$V$ curves in the limit (i) are depicted in Figs.~\ref{fig:on} (thin lines),
\ref{fig:off}b, and \ref{fig:exact-vs-approx}. 
They show a monotonically increasing current, which 
exhibits a step 
at $e V_{sd} \simeq 2\varepsilon_g$ of width $\delta V_{sd}$ increasing with $\tau$ 
and rapidly saturates at an $\varepsilon_g$-independent value 
$I_{sd}^{s}=\pi e\Gamma_0/h$. Such curves are usually shown in 
textbooks, and this feeds the lore of the absent NDR in uncorrelated systems.

What is wrong with the naive argument against the NDR in uncorrelated systems
is that the transmission is \emph{not} independent of $V_{sd}$.
The $V_{sd}$-dependence enters via 
the electrode densities of states $\Gamma_{S,D}$ [cf.~Eq.~(\ref{eq-Sigma-x})].

On one side, this dependence is considered by the $\theta$ functions of 
Eq.~(\ref{eq-Sigma-x}), which diminish the window of allowed tunneling 
processes.
Approximation (ii) that accounts for this yields two qualitatively correct results: 
an NDR beyond $V_{sd}^{\ast}$, where the predicted $I$-$V$ curve
exhibits a cusp (Fig.~\ref{fig:fb}) and a vanishing current for $e V_{sd} \geq 4t$. Quantitatively, 
the NDR onset (at $V_{sd} = V_{sd}^{\ast}$) is unsatisfactory; compare these approximate curves
(label $fb$) with the exact ones in Figs.~\ref{fig:exact-vs-approx} and \ref{fig:fb}.
The NDR occurs well below the point predicted by this approximation.

On the other side, not only the $\theta$ functions, but also the $\varepsilon$-dependence
of the electrode DOS
[the square roots in Eq.~(\ref{eq-Sigma-x})] is important.
It is this fact that makes the finite bandwidth argument incomplete. 
The $\varepsilon$-dependence of $\Gamma_{S,D}$ is accounted for within approximation (iii).
The comparison with the exact curves (Fig.~\ref{fig:exact-vs-approx}) reveals an excellent agreement 
at sufficiently higher voltages (as assumed within this approximation) and demonstrates that, to describe 
quantitatively the NDR, one has to consider both the allowed energy window, which is finite, 
and the energy dependence of the electrode DOS.

\begin{figure}[htb]
% $ $\\[7ex]
% \centerline{\hspace*{-0ex}\includegraphics[width=0.3\textwidth,angle=0]{fig_V_crossover_various_ed.eps}}
\centerline{\hspace*{-0ex}\includegraphics[width=0.5\textwidth,angle=0]{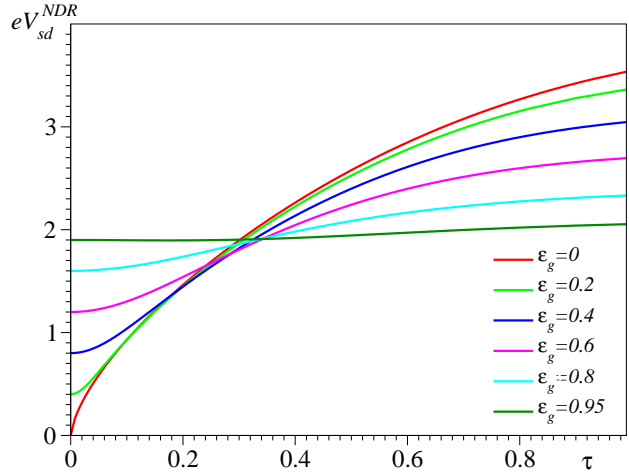}}
\caption{\label{fig:V-cross} (Color online) 
Curves for the NDR onset voltage $V_{sd}^{NDR}$ computed 
from Eq.~(\ref{eq-V-ndr}) for several level energies $\varepsilon_g$. Notice that 
for smaller electrode-dot couplings $\tau$ and not too far away from resonance, 
$V_{sd}^{NDR}$ is significantly smaller than 2 (half of electrode's bandwidth).}
\end{figure}

In Fig.~\ref{fig:exact-vs-approx}, we present exact $I$-$V$ characteristics
from Eq.~(\ref{eq-I}) along with those computed within the three aforementioned approximations, 
Eqs.~(\ref{eq-wbl}), (\ref{eq-fb}), and (\ref{eq-be}). 
%%%%%%%%%%%%%%%%%%%%%%%%%%%%%%%%%%%%%%%%%%%%%%%%%%%%%%%%%%%%%%%%%%%%%%%
%%%%%%%%%%%%%%%%%%%%%%%%%%%%%%%%%%%%%%%%%%%%%%%%%%%%%%%%%%%%%%%%%%%%%%
As visible there, approximation (i) 
is accurate for lower voltages, while approximation (iii) is accurate 
for higher voltages. 
The crossover occurs at a voltage $V_{sd}^{NDR}$, which can be identified with the 
NDR onset. This value can be obtained by equating 
%%%%%%%%%%%%%%%%%%%%%%%%%%%%%%%%%%%%%%%%%%%%%%%%%%%%%%%%%%%%
\begin{equation}
\label{eq-V-ndr}
I_{sd}^{low}(V_{sd}^{NDR}) = I_{sd}^{high}(V_{sd}^{NDR}) . 
\end{equation}
%%%%%%%%%%%%%%%%%%%%%%%%%%%%%%%%%%%%%%%%%%%%%%%%%%%%%%%%%%%%%%%%%%%%%%%
Curves for 
$V_{sd}^{NDR}$ are presented in Fig.~\ref{fig:V-cross}. They show that 
for situations not very far away from resonance and sufficiently weak electrode-dot 
couplings $\tau$, 
$V_{sd}^{NDR}$ is considerably smaller than 
the value $e V_{sd}^{\ast} = 2 t$ expected from the finite bandwidth argument. 
The significant departure of the NDR onset predicted exactly and within approximation (ii) 
is also clearly depicted in Fig.~\ref{fig:fb}. For smaller 
$\tau$'s one can deduce an analytical estimate ($c\simeq 4$)
%%%%%%%%%%%%%%%%%%%%%%%%%%%%%%%%%%%%%%%%%%%%%%%%%%%%%%%%%%%%
\begin{equation}
\label{eq-ndr-onset}
V_{sd}^{NDR} \simeq 2\varepsilon_g + c (t\tau^2)^{1/3} .
\end{equation}
%%%%%%%%%%%%%%%%%%%%%%%%%%%%%%%%%%%%%%%%%%%%%%%%%%%%%%%%%%%%%%%%%%%%%%%
%%%%%%%%%%%%%%%%%%%%%%%%%%%%%%%%%%%%%%%%%%%%%%%%%%%%%%%%%%%%%%%%%%%%%%%
%%%%%%%%%%%%%%%%%%%%%%%%%%%%%%%%%%%%%%%%%%%%%%%%%%%%%%%%%%%%%%%%%%%%%%

%%%%%%%%%%%%%%%%%%%%%%%%%%%%%%%%%%%%%%%%%%%%%%%%%%%%%%%%%%%%%%%%%%%%%%%
%%%%%%%%%%%%%%%%%%%%%%%%%%%%%%%%%%%%%%%%%%%%%%%%%%%%%%%%%%%%%%%%%%%%%%

Interesting for nanotransport are the electron level(s) not too misaligned 
with 
electrode's Fermi level; otherwise, as illustrated by the 
curve for $\varepsilon_g = t$ in Fig.~\ref{fig:off}a, 
the current is very small. Therefore,
the results on $V_{sd}^{NDR}$ expressed by Eq.~(\ref{eq-ndr-onset}) 
and Fig.~\ref{fig:V-cross} are perhaps the most relevant ones 
from an experimental perspective. 
At resonance and realistic parameters 
($t\simeq 1$\,eV, $\tau \simeq 1$\,meV \cite{Goldhaber-GordonNature:98}), 
Eq.~(\ref{eq-ndr-onset}) yields $V_{sd}^{NDR} \simeq 40$\,meV.
Based on this estimate, we argue that the NDR discussed here can be observed. 
On one side, correlations are important only at much lower voltages;
in single-electron transistors,\cite{Goldhaber-GordonNature:98}
the relevant scale is the Kondo temperature $T_K$ 
($e V_{sd} \alt k_BT_K \alt 0.1$\,meV). 
For voltages of tens of mV, correlation effects (e.~g., Kondo's) are supprressed; 
the present uncorrelated limit is justifiable. 
On the other side, the estimated NDR onset voltages ($\sim 10$\,mV) 
are much lower than the electrode bandwidth ($\sim 1$\,eV),
and a material damage prior to the NDR onset can be ruled out. 
For Si-based SETs, the material can support 
even much higher values, $V_{sd}\sim 1$\,V.\cite{Fujiwara}
So, we hope that the present estimate will stimulate experimentalists to search 
NDR effects at moderate $V_{sd}$.
Again quite relevant for experiments, the NDR onset can be controlled by tuning 
the level's energy $\varepsilon_g$ with the aid of a gate potential. 
Gating methods were routinely employed for nanosystems 
in the past\cite{Goldhaber-GordonNature:98} and recently also
in molecular transport.\cite{Reed:09}
In (weakly-correlated) molecules, the level $\varepsilon_g$ would be either the 
highest occupied molecular orbital 
(HOMO)\cite{Reed:09} or the lowest unoccupied molecular orbital 
(LUMO, as in Fig.~\ref{fig:setup}), depending on which is closer to $\varepsilon_F$.
There, $\tau \sim 1$\,eV and $\vert\varepsilon_g\vert \sim 1$\,eV.\cite{Reed:09}
So, the NDR-onset [cf.~Eq.~(\ref{eq-V-ndr}) and Fig.~\ref{fig:V-cross}] is expected at
$V_{sd}$-values of a few eV, slightly higher than used in experiment.\cite{Reed:09}

The present analysis can be extended 
without difficulty to nanosystems/molecules with several ``active'' electron levels.
As long as these levels $\varepsilon_{g 1}, \varepsilon_{g 2},\ldots$ 
are well separated energetically and the hybridization is weak enough 
(a different situation can also be encountered, see Ref.~\onlinecite{Peskin:09}),
they manifest themselves as current steps at the voltages 
$e V_{sd} \approx 2 \varepsilon_{g 1}, 2 \varepsilon_{g 2}, \ldots$.
However, even in this case the finite electrode bandwidth and the energy dependence 
of the electrode DOS remain possible important sources of an NDR.

Similar to 
other situations encountered in nanotransport,\cite{Baldea:2008b,Baldea:2009c}
we believe that the results for uncorrelated systems are instructive and 
could also be useful 
to correctly interpret the nanotransport 
in correlated systems. In the present concrete case, 
they could help to unravel the physical origin of the NDR. 
In the light of the present analysis,
it is plausible to ascribe an NDR as an electron correlation effect in cases 
where the NDR was found within  
calculations to a correlated nanosystem carried out within the wide band limit. 
This is, e.~g., the case 
of Refs.~\onlinecite{Doyon:07} and \onlinecite{Nishino:09}, 
where a weaker NDR effect was obtained at resonance 
at stronger Coulomb contact interactions. As suggested by Fig.~\ref{fig:off}, 
the farther away from resonance, the more is the NDR onset pushed towards 
higher voltages ($e V_{sd}^{NDR} > 2 \vert\varepsilon_g\vert$).
The values of $V_{sd}$ chosen in the figures shown in  
Ref.~\onlinecite{Mehta:07} do not belong to this range and the absence 
of an NDR could be related to this fact. 
Unlike the wide (infinite) band limit assumed in the aforementioned 
references, a discrete model of the electrodes, with a finite bandwidth $4 t$, 
\emph{exactly} as in
Eq.~(\ref{eq-ham}), has been utilized for the numerical calculations of 
Ref.~\onlinecite{Schmitteckert:08} at resonance. The $I$-$V$ curves reported 
there exhibit a pronounced NDR effect. 
However, in view of the finite bandwidth 
assumed in that work, attributing this effect to electron correlations 
at rather high voltages should be 
made with special care. We believe that in order to interpret this effect reliably, 
one should first carefully subtract
the contribution to the NDR due to the finite bandwidth and the energy dependent 
electrode DOS discussed above.

The financial support for this work  
from the Deu\-tsche For\-schungs\-ge\-mein\-schaft is gratefully acknowledged.
%%%%%%%%%%%%%%%%%%%%%%%%%%%%%%%%%%%%%%%%%%%%%%%%%%%%%%%%%%%%%%%%%%%%%%
% \bibliographystyle{aip}
% \bibliographystyle{apsrev}
% \bibliography{/home/ioan/QDs/LinearResponse/paper/bibl}% Produces the bibliography via BibTeX.
% \end{document}
%%%%%%%%%%%%%%%%%%%%%%%%%%%%%%%%%%%%%%%%%%%%%%%%%%%%%%%%%%%%%%%%%%%%%%

\end{document}